\documentclass[seceq,amsmath,amssymb]{ptptex}

\usepackage{graphicx}
\usepackage{bm}

\def\bea{\begin{eqnarray}}
\def\eea{\end{eqnarray}}

\def\pp{\mbox{$p$-$p$} }
\def\auau{\mbox{Au-Au} }
\def\cucu{\mbox{Cu-Cu} }

\def\aa{\mbox{A-A} }
\def\nn{\mbox{N-N} }

\title{
The ``soft ridge'' -- is it initial-state geometry or modified jets?
}

\author{
Thomas A. \textsc{Trainor}%
}

\inst{
CENPA 354290, University of Washington, Seattle, WA 98195 USA
}

\abst{
An $\eta$-elongated same-side 2D peak (``soft ridge'') in minimum-bias angular correlations from heavy ion collisions has been attributed both to jet formation and to initial-state geometry structure coupled to radial flow. We consider evidence for both interpretations.
}

\begin{document} 
\maketitle

 \section{Introduction}

A same-side (SS) 2D peak dominates minimum-bias angular correlations (no trigger-associated $p_t$ cuts) for all \aa centralities at higher RHIC energies. In 200 GeV \pp collisions the SS peak properties are consistent with minijets. The peak is elongated on $\phi$.~\cite{porter} In more-central \auau collisions the SS peak becomes elongated on $\eta$~\cite{anomalous} and is then described by some as a ``soft ridge.''  Recent initiatives reinterpret the ``soft ridge'' in terms of flows.~\cite{gmm,gunther,sorensen}
A recipe for assigning the SS 2D peak to (higher harmonic) flows has emerged: (a) Project (all or part of) the $\eta$ acceptance onto azimuth $\phi$; (b) fit the 1D projection on $\phi$ with a Fourier series; (c) interpret each series term as a "harmonic flow;" (d) attribute the flows to conjectured A-A initial-state (IS) geometry. To better establish the true mechanism for the SS 2D peak we compare recent flow conjectures and a minijet interpretation within a 2D context.

 \section{Angular correlations from minimum-bias jets (minijets)}

Substantial experimental and theoretical evidence supports the conclusion that SS 2D and away-side (AS) 1D peaks are manifestations of minimum-bias jets (minijets).~\cite{porter,anomalous,fragevo,jetsyield} The monolithic minimum-bias SS 2D peak is well described by a 2D Gaussian with no additional ridge structure. The AS 1D peak is consistent with parton momentum conservation (dijets). AS correlation structure is uniform on $\eta$.

 \begin{figure}[h]
 \hfill \includegraphics[width=1.65in,height=1.4in]{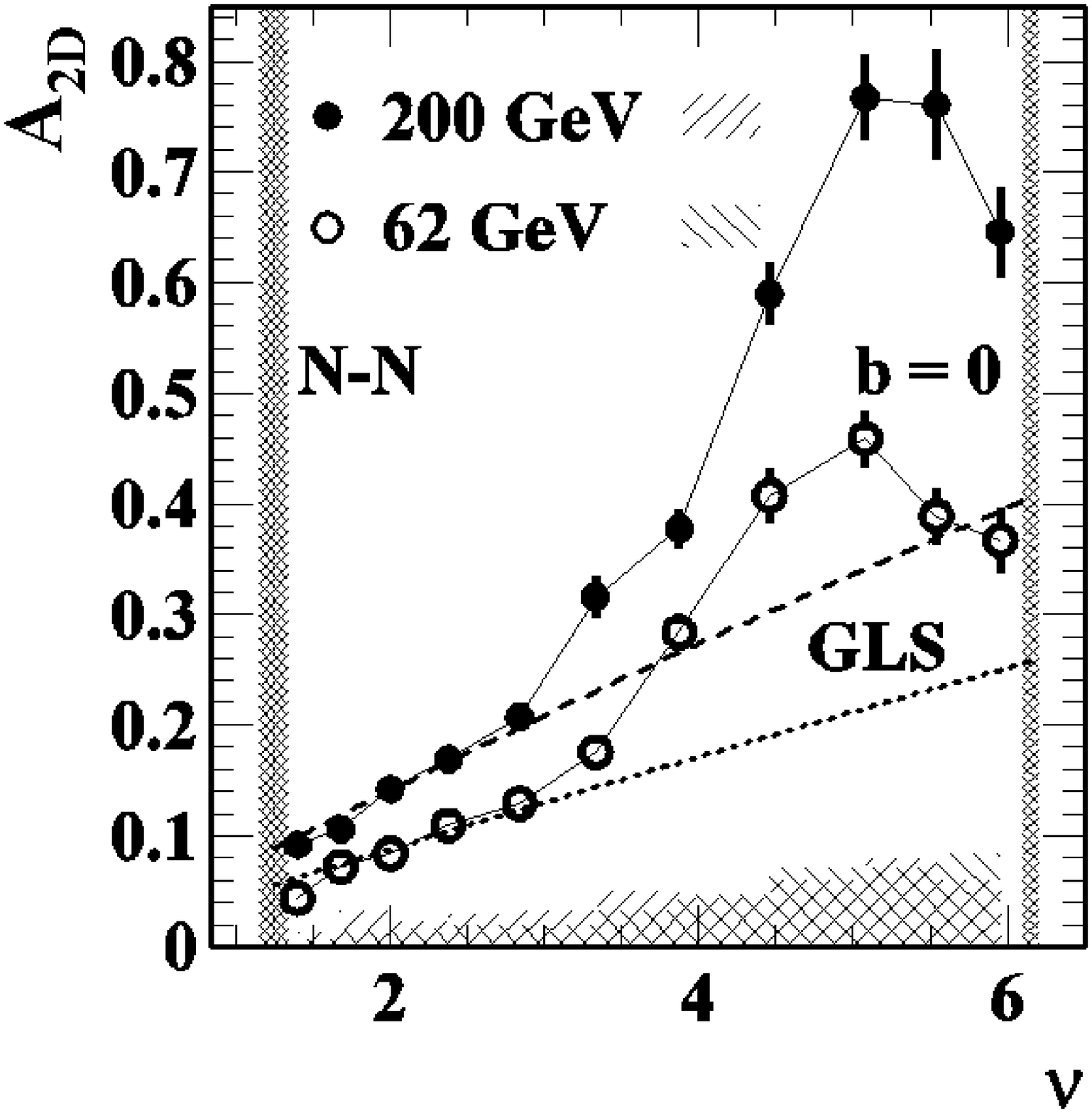}
  \includegraphics[width=1.65in,height=1.42in]{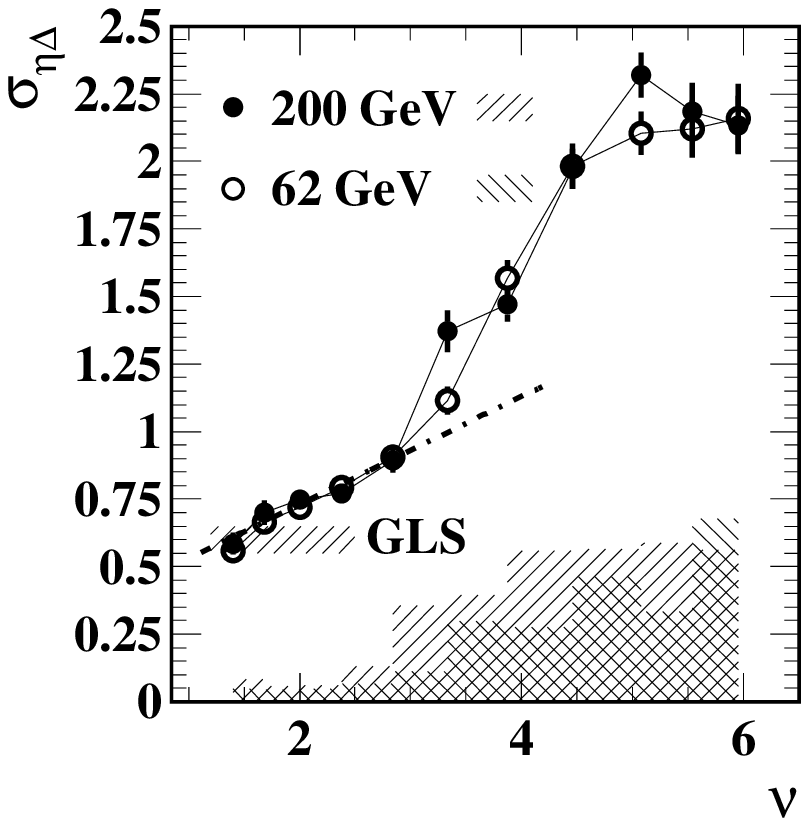}
  \includegraphics[width=1.65in,height=1.42in]{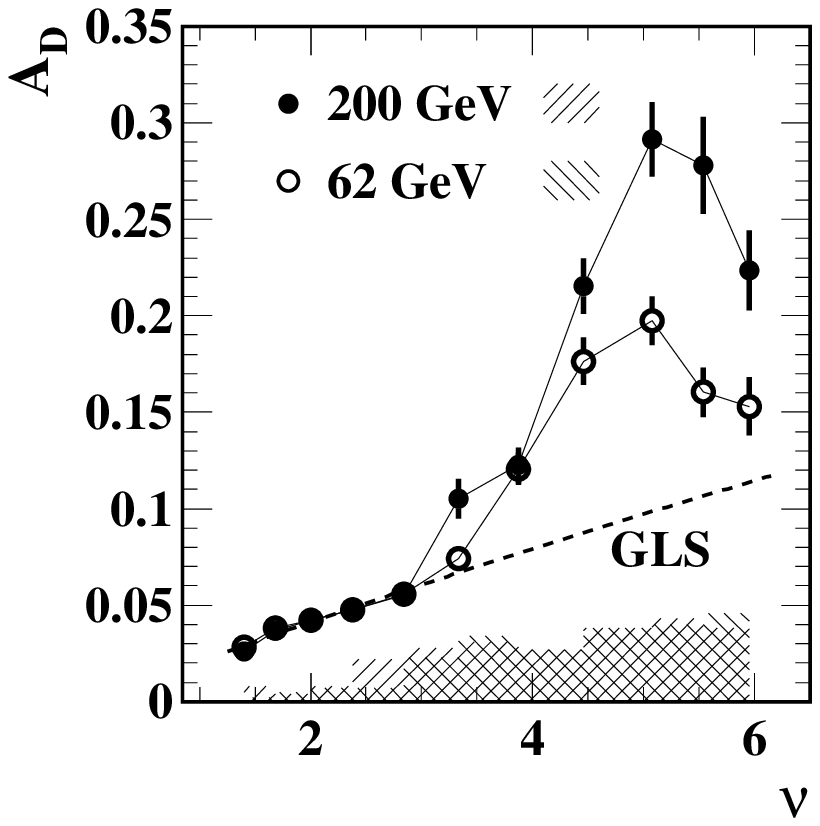}
\hfill
\hfill
\caption{\label{ssparams}
First: SS 2D peak amplitude,
Second: SS peak $\eta$ width,
Third:  AS 1D peak amplitude.
}
 \end{figure}

Figure~\ref{ssparams} shows \auau centrality trends for SS 2D and AS 1D peak parameters.~\cite{anomalous} The trends are consistent with minijets from Glauber linear superposition (GLS) of \nn collisions up to a sharp transition at $\nu \equiv 2N_{bin} / N_{part} \approx 3$. Above that point deviations from GLS are consistent with modified parton fragmentation to jets.~\cite{hardspec,fragevo}


 \section{SS 2D peak Fourier decomposition and ``higher harmonics''}

\begin{figure}[h]
 \hfill \includegraphics[width=1.65in,height=1.4in]{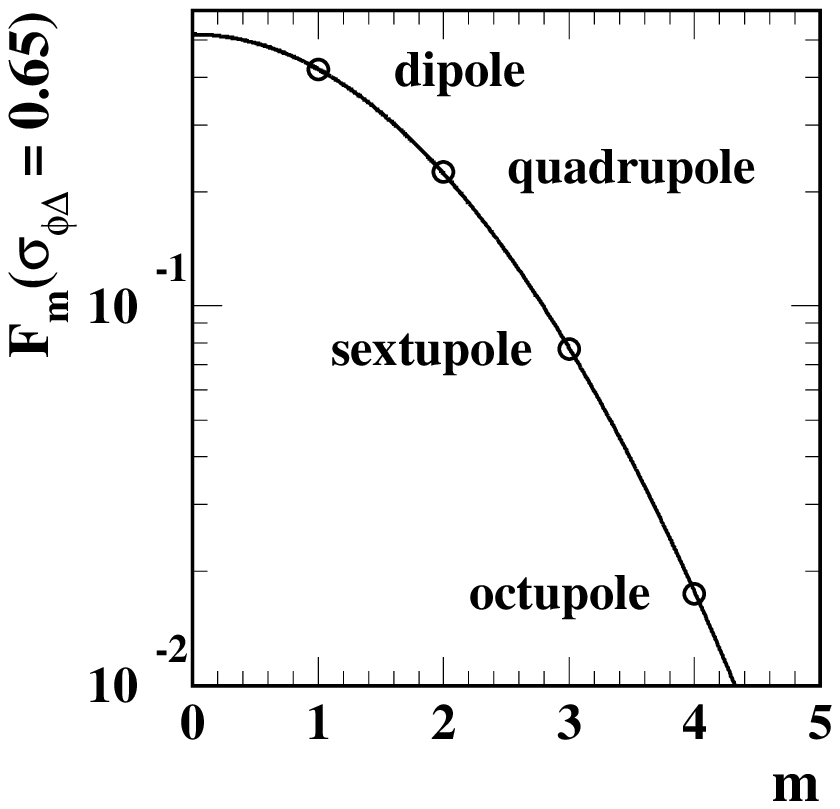}
  \includegraphics[width=1.65in,height=1.4in]{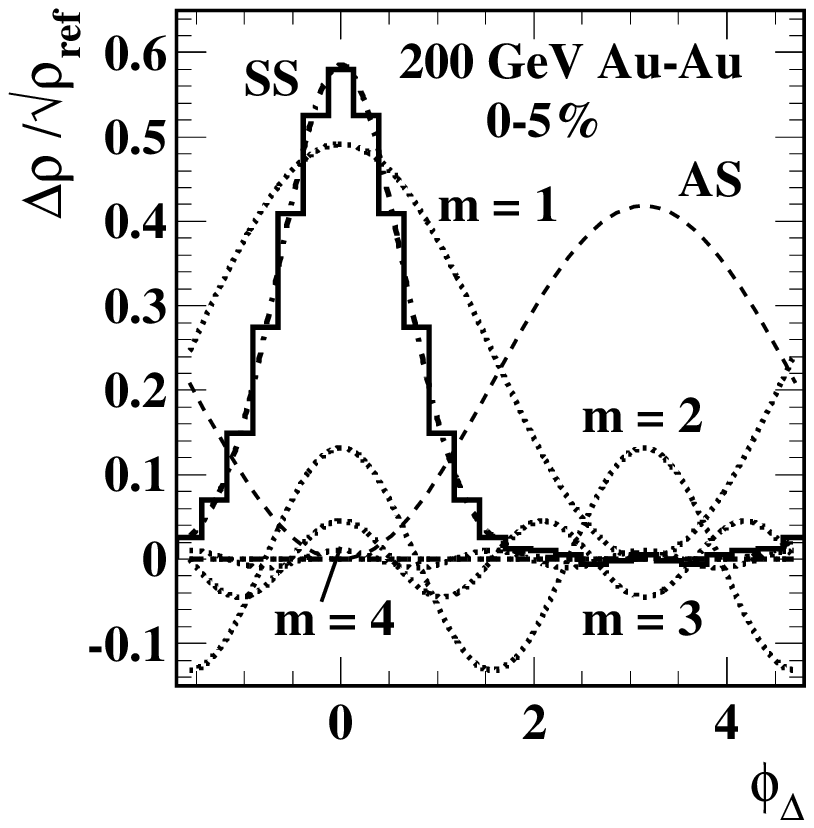}
  \includegraphics[width=1.65in,height=1.4in]{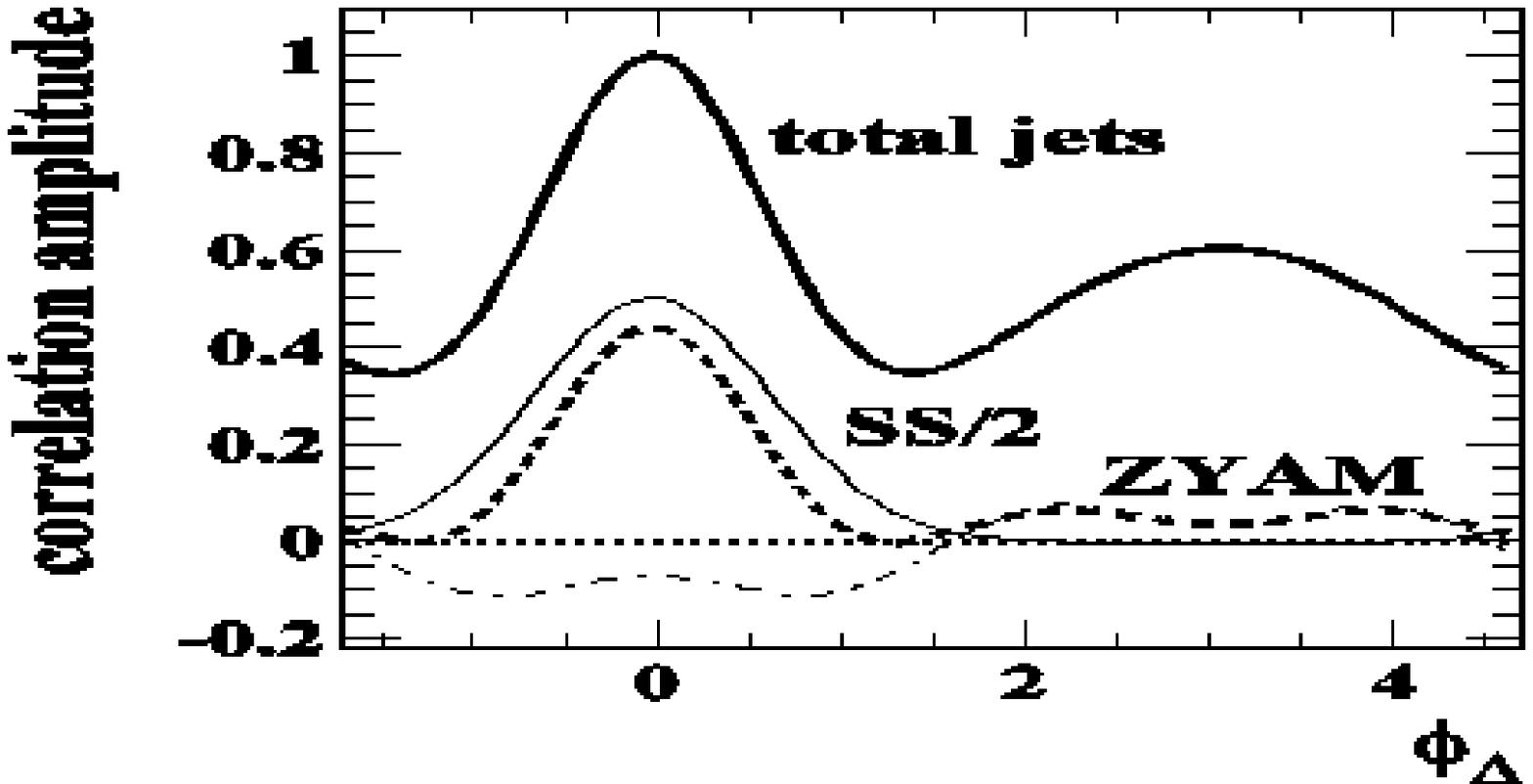}
\hfill
\hfill
\caption{\label{fourier}
First: Fourier coefficients, 
Second: SS peak Fourier components,
Third: ZYAM subtraction.
}
 \end{figure}

If the SS 2D peak is projected onto 1D azimuth the resulting 1D Gaussian has a Fourier series representation, the terms representing cylindrical multipoles. The jet-related quadrupole amplitude is $2A_Q\{SS\} = 2\rho_0(b) v_2^2\{SS\} = F_2(\sigma_{\phi_\Delta}) A_{1D}$, where $A_{1D}$ is the projected SS 1D peak amplitude, $F_2(\sigma_{\phi_\Delta})$ is a Fourier coefficient and $\rho_0(b)$ is the single-particle angular density. 
Figure~\ref{fourier} (first panel) shows Fourier coefficients for a unit-amplitude Gaussian with r.m.s.\ width $\sigma_{\phi_\Delta} = 0.65$ (SS 1D peak width for more-central 200 GeV \auau collisions). The coefficients are given by $F_m(\sigma_{\phi_\Delta}) = \sqrt{2/\pi} \, \sigma_{\phi_\Delta} \exp(-m^2 \sigma^2_{\phi_\Delta} / 2)$. Thus, ``higher harmonics'' from the SS 2D peak can be predicted accurately from measured minijet systematics as in Fig.~\ref{ssparams}.

Figure~\ref{fourier} (second panel) shows the measured SS 2D peak from 0-5\% central \auau collisions projected onto azimuth (bold histogram).~\cite{anomalous} The fitted AS dipole (dashed curve) has been subtracted from the data histogram. RMS residuals from the 2D model fit are $\approx 0.5$\% of the SS peak amplitude, consistent with statistics. The multipole components of the SS 1D peak (dotted curves) are calculated from $F_m$ and the SS peak parameters. For 0-5\% central collisions the nonjet quadrupole  $A_Q\{2D\} \approx 0$.~\cite{davidhq} The $A_X\{SS\}$ ($X = Q,\,S,\,O$) represent structures with large curvatures on $\eta_\Delta$, whereas $A_Q\{2D\}$ represents a structure uniform on $\eta_\Delta$ within the STAR TPC acceptance. The $\eta_\Delta$ dependence permits accurate distinction between jet-related multipoles $A_X\{SS\}$ (``nonflows'') and the nonjet quadrupole $A_Q\{2D\}$.

The {\em total} quadrupole from projected angular correlations is $A_Q\{2\} = A_Q\{SS\} + A_Q\{2D\}$. A similar expression holds for higher moments $Q \rightarrow S, O$ (sextupole and octupole). The corresponding $v_m$ derived from SS 2D peak parameters in Fig. 1 for 0-5\% 200 GeV \auau are $v_2\{2\} = 0.026$, $v_3\{2\} = 0.015$ and $v_4\{2\} = 0.007$. By the same procedure ``triangular flow'' and higher multipoles can be derived from measured SS 2D peak systematics with various $\eta$ exclusion cuts supposed to reduce or eliminate ``nonflow.''~\cite{multipoles} ``Higher harmonic flow'' results from the LHC can be (and have been) anticipated by RHIC minijet and nonjet quadrupole measurements.~\cite{multipoles}

Figure~\ref{fourier} (third panel) illustrates ZYAM subtraction to infer jet structure from ``triggered'' dihadron correlations. The data (plotted relative to a zero offset determined by 2D model fits) are described by the bold solid curve. ZYAM subtraction with conventional $v_2$ estimate from published data leads to the dashed curve: greatly reduced peak amplitudes and AS double peak. The dash-dotted curve is the sextupole from the SS jet peak plus the difference between SS and AS dipoles, explaining the AS  double-peak structure. The jet inference is distorted and misleading.~\cite{tzyam}

 \section{Comparing jet, nonjet quadrupole and IS geometry trends}

Accurate distinction between jets and nonjet structures attributed to conjectured flows depends on careful differential comparisons among $p_t$, $\eta$ and centrality dependence of angular correlations. For example, the nonjet quadrupole $A_Q\{2D\}$ inferred from 2D model fits to angular correlations is accurately distinguished from jet structure on the basis of the strong $\eta$ dependence of the SS 2D peak. 

\begin{figure}[h]
 \hfill 
  \includegraphics[width=1.65in,height=1.42in]{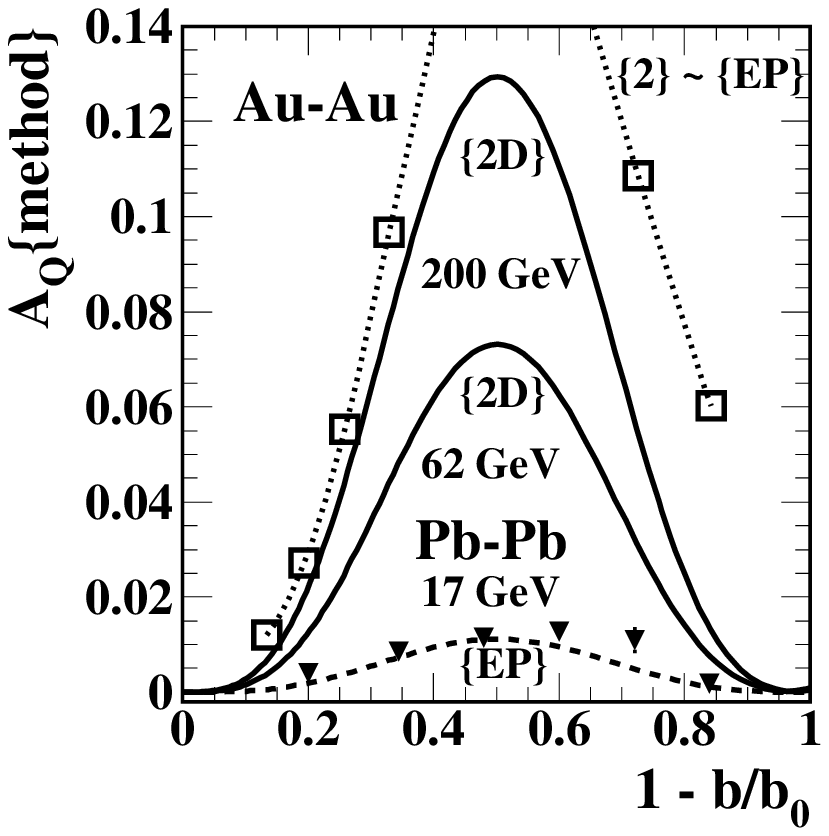}
\includegraphics[width=1.65in,height=1.4in]{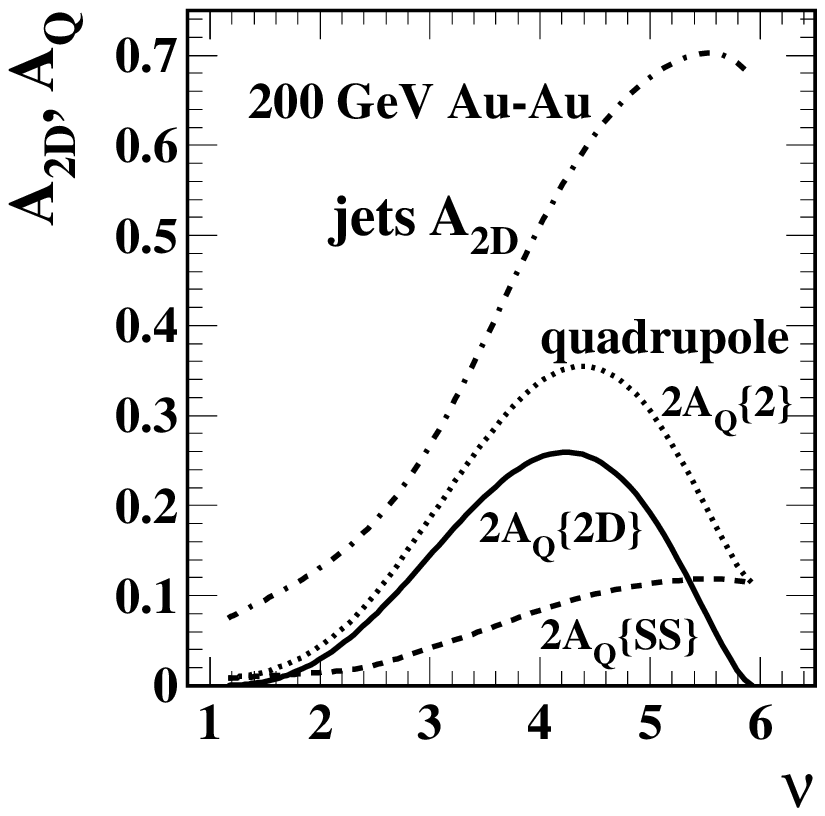}
  \includegraphics[width=1.65in,height=1.42in]{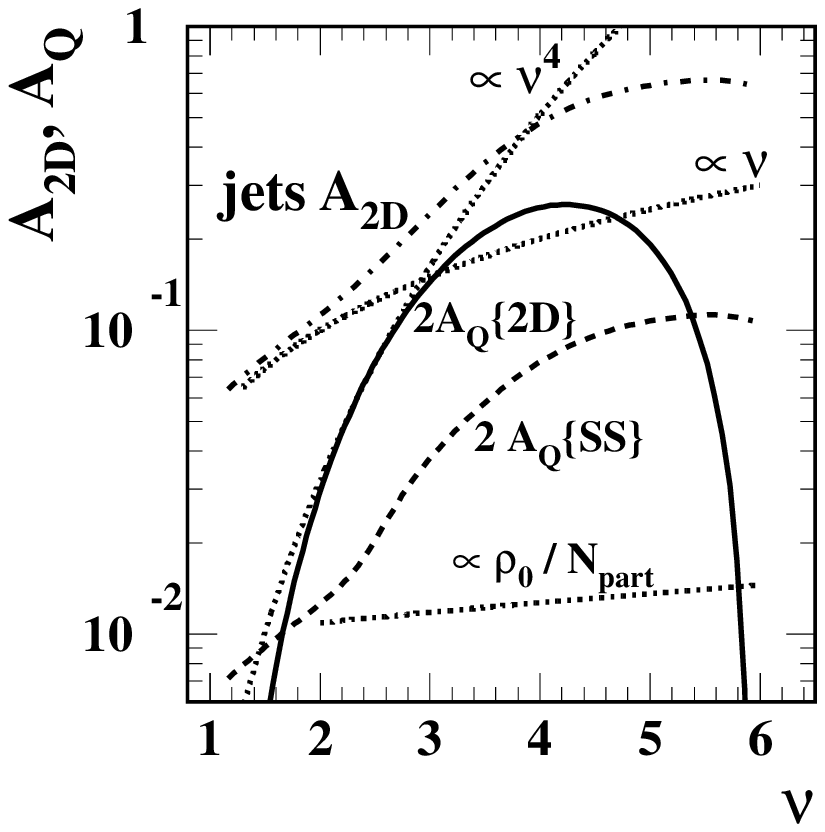}
\hfill
\hfill
\caption{\label{compare}
First: Azimuth quadrupole,
Second: Jet and nonjet trends,
Third: Centrality comparisons.
}
 \end{figure}

Figure~\ref{compare} (first panel) shows $A_Q\{2D\}(b)$ data (solid curves) inferred from 2D model fits.~\cite{davidhq} The dashed curve is the same trend extrapolated to 17 GeV. The open square points and solid triangles represent $A_Q\{EP\} \approx A_Q\{2\}$ transformed from published $v_2\{EP\} \approx v_2\{2\}$ data.~\cite{2004,na49} The $A_Q\{2D\}$ data exhibit universal centrality and energy trends for a phenomenon independent of SS and AS jet structure.

Figure~\ref{compare} (second panel) shows a direct comparison among jet and nonjet structures. The relation between $v_2$ methods is given by $A_Q\{2\} = A_Q\{SS\} + A_Q\{2D\}$: jet-related plus nonjet quadrupoles sum to a total quadrupole amplitude measured by $v_2\{2\} \approx v_2\{EP\}$. ``Nonflow'' component $2A_Q\{SS\} = F_2(\sigma_{\phi_\Delta}) G(\sigma_{\eta_\Delta}) A_{2D}(b)$ is the quadrupole component of the SS 2D peak projected onto azimuth.~\cite{multipoles} The $v_2\{2\}$ inferred from that relation accurately describes published $v_2$ data.~\cite{gluequad,multipoles} 
The distinction between jet and nonjet structure is based on curvatures on $\eta_\Delta$.

Figure~\ref{compare} (third panel) compares centrality trends for several correlation mechanisms. The nonjet quadrupole $A_Q\{2D\}$ (bold solid curve) varies as $N_{bin} \epsilon_{opt}^2 \propto \nu^4 \epsilon_{opt}^2$~\cite{davidhq}, a very strong rate of increase for more-peripheral collisions. The SS 2D peak amplitude varies as $N_{bin} / n_{ch} \approx \nu = 2N_{bin} / N_{part}$ (upper dotted curve) for Glauber linear superposition but increases more rapidly in \auau collisions above a sharp transition at $\nu \approx 3$ (dash-dotted curve).~\cite{anomalous} The corresponding jet-related quadrupole $A_Q\{SS\}$ (dashed curve) includes the effect of projecting the SS 2D peak onto 1D azimuth.~\cite{multipoles} All jet-related higher multipoles (``higher harmonic flows'') share the same centrality trend. The statistically compatible IS geometry measure $\rho_0 \epsilon^2_{m,MC}$ (lower dotted curve) varies as $\rho_0 / N_{part} \approx $ constant for $m$ odd.~\cite{multipoles} If those three elements are actually related to a common IS geometry through hydrodynamic flows why are the centrality trends so dramatically different?

\section{Jets vs flows in a larger context}

The correlation structures attributed separately to minijets and the nonjet quad-rupole in 2D analysis can be compared in other contexts, including particle yields and $p_t$ spectrum structure. Quadrupole $p_t$ spectra can be inferred from $v_2(p_t,b)$ data for unidentified and identified hadrons.~\cite{quadspec} The quadrupole component emerges from a boosted source with fixed boost independent of \aa centrality, and the quadrupole appears to be carried by a small fraction of the total hadron yield.~\cite{davidhq2} Single-particle yields inferred from minijets plus a pQCD jet cross section agree with inferred spectrum hard-component yields.~\cite{hardspec,jetsyield} 
Spectrum hard-component systematics in turn agree with pQCD calculated parton fragment distribution.~\cite{fragevo} Glasma flux tubes as a mechanism for the SS 2D peak disagree with the $p_t$ structure of correlations.~\cite{glasma}

\section{Summary}

Minimum-bias 2D angular correlations include a monolithic same-side 2D peak, an away-side 1D peak described by a single dipole shape and a nonjet azimuth quadrupole represented by $v_2\{2D\}$.  The SS 2D peak and AS 1D peak are quantitatively related to minimum-bias pQCD jets.
Conventional $v_m$ analysis ignores the $\eta$ structure of the SS 2D peak. Consequently, Fourier components of the SS peak bias all $v_m\{2\}$ data as ``nonflows.''  Recently announced ``higher harmonic flows'' are Fourier components of the SS peak.  ZYAM subtraction of the jet-related quadrupole and  higher harmonics is equivalent to subtracting  jets from jets. In effect, parton fragmentation scenarios in nuclear collisions have been abandoned without regard to likely  jet modifications (e.g., $\eta$ broadening) in the A-A collision environment.


\end{document}